CHAPTER 4

# China, India, and Myanmar: Playing Rohingya Roulette?

*Hossain Ahmed Taufiq*

## Introduction

It is no secret that both China and India compete for superpower standing in the Asian continent and beyond. Both consider South Asia and Southeast Asia as their power-play pivots. Myanmar, which lies between these two Asian giants, displays the same strategic importance for China and India, geopolitically and geoeconomically. Interestingly, however, both countries can be found on the same page when it comes to the *Rohingya* crisis in Myanmar's Rakhine state.

As the Myanmar army (the *Tatmadaw*) crackdown pushed more than 600,000 *Rohingya* refugees into Bangladesh, Nobel Peace Prize winner Aung San Suu Kyi's government was vociferously denounced by the Western and Islamic countries.[1] By contrast, China and India strongly supported her beleaguered military-backed government, even as Bangladesh, a country both invest in heavily, particularly on a competitive basis, has sought each to soften Myanmar's *Rohingya* crackdown and ease a mediated refugee solution.

H. A. Taufiq (✉)
Global Studies & Governance Program, Independent University of Bangladesh, Dhaka, Bangladesh
e-mail: taufiq@iub.edu.bd

81

China's and India's support for Myanmar is nothing new. Since the Myanmar military seized power in September 1988, both the Asian powers endeavoured to expand their influence in the reconfigured Myanmar to protect their national interests, including heavy investments in Myanmar, particularly in the Rakhine state. Though their policies towards Myanmar had many similarities, as well as significant differences in content and results, the unintended consequences of Bangladesh paying the costs may tilt Bangladesh one way or the other.

## Chinese Strategic Interests and Stakes in Myanmar

China and Myanmar share a close linguistic and anthropologic link. With both Burmese and Chinese people being of Sino-Tibetan origin, despite their close connections, only during the perilous World War II did China realise how strategically significant Burma (Myanmar's other name) was. Famous as it is, the "Burma Road," for example, illustrates China-British Burma cooperation during that time, playing, in fact, a critical role in defeating Imperial Japan. Originating in India's Assamese footholds, Burma Road proved as a lifeline to the Chinese, British, and allied forces. They used it as a safe passage for moving weapons and possessions into Japanese-besieged China and Burma.[2] Myanmar started to play a crucial role in China's foreign policy calculations ever since the People's Republic of China was founded in 1949. China has several policy objectives in Myanmar, including access to the Indian Ocean, energy security, border stability, and bilateral economic cooperation.[3] China's foremost urgency to develop peacefully, maintaining its political, security and strategic, and economic objectives, indicates its desire in Myanmar. In other words, China's search for a gateway to the Indian Oceanic resulted in its multidimensional policies towards Myanmar.[4] It did not emerge suddenly, but since years of change and refinements shaped it to what it is now, a closer look helps our understanding.

### *Access to the Indian Ocean*

Myanmar shares a 2000 km long border with China's Yunan province, which China considers the bridgehead to Southeast Asia, South Asia, and two oceans—the Indian Ocean and the Pacific. Yet, the U.S. control of the adjacent Strait of Malacca has long irritated China, since it is the crucial passageway of the oil it imports, symbolically as strategically significant to its security as the Strait of Hormuz was to the United States, indeed, West Europe, and Japan until recently. For decades, China sought an alternate trade route to Africa, Asia, and the Middle East. Since Myanmar offers that opportunity, China has invested upwards of USD 15 billion in Myanmar under the sponsorship of its "The Great International Yunnan Passage."[5] This route comprises of a comprehensive set of road, rail, and air connections as well as water, oil, and gas pipelines. China believes this investment will help uplift its southwestern provinces. According to Voon Phin Keong, the director of the Centre of Malaysian Chinese Studies in Kuala Lumpur, Myanmar is "an outlet … would enable China to overcome its 'single-ocean strategy' and to realise what would constitute a highly significant plan for a 'two-ocean strategy.'"[6] Since Chinese scholars and officials had for long pressed the government to adopt an Indian Ocean strategy and build international channels to the Indian Ocean, the development of the Great Yuan passage is the answer to their desire.[7]

With the Kunming-Ruili-Yangon and Kunming-Tengchong-Myanmar-India roads being the updated versions of the famous Burma Road, one off-shoot, the Kunming-Ruili-Muse-Mandalay-Yangon, has become the leading China-Myanmar trade and connectivity route.[8]

The Yunnan government has spent millions in modernising key trading infrastructures bordering Myanmar, such as the Ruili Development and Experimental Zone (RDEZ).[9] It is the most significant land port of Yunnan. RDEZ is linked with Muse, a border town and port in Myanmar's northern Shan state. Muse is the passage to Myanmar's second largest city, Mandalay, via Lashio. Burma Road also passes through Mandalay, Lashio, and Muse, before it connects with Yunan. Also, the "Great passage" plan includes a railroad from Yunan to Kyauk Phyu of the Rakhine state.[10]

*Energy Security*

In 1993, China emerged as a net oil importer, but by now it is one of the most energy-hungry countries in the world, with a dependency reaching 64.4 per cent of total oil consumption in 2017 (up from 29 per cent in 2000).[11] Further growth is expected.[12] About 80 per cent of China's imported oil passes through the Malacca Strait, making it highly vulnerable, for example, even to an embargo, let alone conflict.[13] Chinese scholars have also long advocated importing oil and gas from the Middle East and Africa through pipelines girding Yunan province to the Kyauk Phyu port.[14] Their two-step plan to avoid the Malacca Strait includes, first, to construct a deep-sea port in Myanmar, where the oil and gas would be brought by ships (in liquefied format), and second, transfer the oil and gas through separate pipelines to southwest China.

China, therefore, has invested heavily in the oil and gas infrastructure development in Rakhine. A 2.45 billion USD pipeline from Western China to Kyauk Phyu is already operational, with the gas pipeline stretching 793 km and the oil pipeline 771 km.[15] These pipelines will provide a secure route for Beijing's imported Middle East crude oil, and significantly reduces China's reliance on oil supplies passing through the U.S.-controlled Strait of Malacca. The oil pipeline can transfer 22 million tons oil per year which is 5–6 per cent of China's net oil import.[16]

A hydrocarbon-rich country, Myanmar also has proven natural gas reserves of about 2.5 trillion cubic metres.[17] These proven reserves equal China's own proven natural gas reserves.[18] China recently received a contract from Myanmar government to explore and develop the gas fields.

### *Economic Development and Cooperation*

Under China's "Belt and Road Initiative," Beijing also has an ambitious plan to invest in the infrastructure sector of the Rakhine state. Worth 7.3 billion USD, the investment package includes the development of Kyauk Phyu Special Economic Zone and a deep seaport in the Bay of Bengal.[19] Although the port project has been jointly awarded to a consortium led by China's CITIC Group Corporation Ltd (formerly the China International Trust Investment Corporation), negotiations continue between Myanmar and China over allocations: the consortium seeks 70–85 per cent, the Myanmar government equal partnership.[20] The port is part of two projects, where the other one is the establishment of an exclusive economic zone, with the CITIC leadership role in both decided in 2015.[21]

The striking resemblance with Sri Lanka's Hambantota Port construction adjacent to an economic development zone, funded by China, should not be missed, especially since China also has a similar structure in Pakistan's Gwadar Port. When Sri Lanka failed to pay back the loan that built these projects, it had to lease the port for 99 years to the Chinese company that built it, along with the economic zone (for a reduced timespan).[22] Myanmar's debt to China is growing, carrying enormous consequences for the country, region, and the world. Viewing these cases in its rear-view mirror cannot but shape Myanmar's continued relationship with China.

### *Military Trade and Cooperation*

Southwestern China security forms a significant component of the Beijing-Naypyidaw relationship. Myanmar receives most of its military hardware and training from the People's Liberation Army of China, but, to avoid any international criticism, China kept its arms trade with Naypyidaw limited—only to large-scale military hardware supplies, not small arms.[23] Such a policy maintains the status quo between the *Tatmadaw* and the ethnic nationalities along the Sino-Myanmar border. Recent procurements by *Tatmadaw* from China include the purchase of HN-5 shoulder-launched missiles for air defence and C-801 and C-802 anti-ship cruise missiles for warships.[24]

## India's Strategic Interests and Stakes in Myanmar

India and Myanmar have a historical connection. Both were part of the extended British Empire in Asia. Relations remained mostly friendly between the two neighbours since both became independent after World War II. Prime ministers Jawaharlal Nehru and U Nu stressed on joint economic developments and worked closely to outline a plan for the future. India even provided limited military support to Myanmar. Moreover, both countries decided to remain neutral in the Cold War by becoming active members of the Non-Aligned Movement (NAM). Relations strained between the two countries after the 1962 military coup. India strongly opposed the military takeover of General Ne Win and supported the pro-democracy forces in Myanmar. Ne Win's anti-Soviet foreign policy at a time when relations between the Soviet Union and India were burgeoning did not help, added to which Myanmar declined to join the Commonwealth of Nations and pulled out from the Non-Aligned Movement in 1979.[25]

Though relations improved from 1988, tensions remained. India kept continuing its sympathy and support for the pro-democracy groups in Myanmar, and, as a demonstration of that support, it awarded Aung San Suu Kyi the Jawaharlal Nehru Award for International Understanding in 1993.[26] The Than Shwe government declared Suu Kyi a persona non grata, but the State Law and Order Restoration Council's isolationist strategy further hampered the renewal of full Indo-Myanmar relations. Relaxation continued again from 1994, as "India adopted a more pragmatic and less moralistic stance,"[27] increasingly more routine politically and diplomatically. Perhaps the most drastic change dates back to November 2001, when General Maung Aye, Vice Chair of the Myanmar State Peace and Development Council, made an ice-breaking visit to India.

India and Myanmar concur on many issues. Frequent reciprocal visits by high-level leaders took place between India and Shwe's isolationist regime.[28] India, at one point, withdrew its support for the National League for Democracy, and in 2006 stopped welcoming the exiled dissidents. The Indian defence minister, on a Singapore visit, reaffirmed India's peaceful coexistence policy approach with Myanmar. Aung Sun Suu Kyi's election-winning in 2012 and 2015, and her ascension to power in 2015, have brightened India-Myanmar relationship. India now sees it as a golden opportunity for forging economic relations.

Military and security concerns form this bilateral relationship. More than 20 exchange visits between military officers from the two countries took place since Myanmar joined the ASEAN (Association of Southeast Asian Nations) group in 1997.[29] At least four telephone hotlines have been established to keep the flawless connection between the military commands of the two countries.[30] Both countries believe vulnerability to non-traditional actor threats can be overcome only by cooperation. During Than Shwe's first visit to India, in October 2014, several agreements were signed, notably a memorandum of understanding underpinning the border security, counter-terrorism, and exchange of security personnel issues.[31]

As bilateral relations improved, India made the tectonic shift to re-launching its 1991 "Look East" policy, now dubbing it "Act East" from 2014 to reiterate the urgency.[32] India considers Myanmar a key ally in maintaining Northeast India NEI (North East India) security and stability. The four NEI states—Mizoram, Manipur, Arunachal Pradesh, and Nagaland—share a common border of 1643 km with Myanmar.[33] India also believes that Myanmar's cooperation is critical to limiting the influence of *Naga* insurgency.[34] Both countries, along with Bangladesh, have also discussed gas-pipelines interconnections, based on new discoveries in the region.

Another key reason for New Delhi's new "Act East" policy is to counter China's growing clout in Southeast Asia and the Bay of Bengal. The convergence of all these aspects has impelled New Delhi to invest heavily in Myanmar, setting forth several maritime and land-based Myanmar infrastructural development plans, such as the landmark Kaladan multimodal project, India-Myanmar-Thailand Asian Trilateral Highway, and a road-river-port cargo transport project.[35]

### *Infrastructure*

By 2001, India had built the 160 km India-Myanmar Friendship Road, which is now included in the India-Myanmar-Thailand Highway Project.[36] Not only that, in the following year, India started the open-sea route to connect Mizoram state in Northwestern India.[37] It also has an agreement with Myanmar to construct the Thahtay Chaung hydropower project in Rakhine state, and joint ventures to produce a 1200-megawatt hydropower dam at Thamanthi in the river basin of Chindwin river.[38]

Energy security is a source of great concern for most emerging countries. India started working with Myanmar to develop their own hydropower programme to supply much-needed power to India and its neighbouring countries.[39]

The most remarkable investment programme between India and Myanmar is the Kaladan Multi-Modal Transit Transport Project, a 484 million USD project.[40] It includes dredging the river, as well as port and road construction to connect Sittwe, Rakhine, and the Chin hinterland, Mizoram. It would reduce the distance between Kolkata to Aizawl about half from its current distance of 1550 km and would provide the landlocked states of India a significant route to the Bay of Bengal bypassing Bangladesh.[41]

So far, India has completed the Sittwe port. The construction of a river terminal and dredging of the river are currently underway. At the same time, India's infrastructural development assumes setting up an SEZ (Special Economic Zone) at Sittwe, near the Chinese Kyauk Phyu port and SEZ facility.

The project's several sections combine multimodes of transport—a 539 km shipping route from Kolkata (India) to Sittwe (Myanmar), a 158 km riverboat route from Sittwe seaport to Inland Water Terminal (IWT) of Paletwa jetty (Myanmar), a 110 km road route from IWT Paletwa to Indo-Myanmar border, and, finally, a 100 km highway route from Zorinpui (Indo-Myanmar border) to Aizawl-Saiha (Mizoram, India).[42]

Meanwhile, a 1360 km highway project to connect India, Myanmar, and Thailand will start from Moreh in India to Mae Sot in Thailand through Myanmar.[43] The aim is to increase India's presence in Rakhine and thereby get ASEAN access through Myanmar.[44]

### *Obtaining Access to Gas and Oil*

Presently, India is the third largest energy-consuming country in the world. It consumes over 4.1 million barrels of crude oil and 378.06 million cubic metres of gas each day.[45] With limited resources within its geographic boundary, it has to import most of the oil and gas from foreign sources. After the discovery of a natural gas field near Sittwe in 2004, India proposed to buy the gas, with a mid-January 2005 agreement with Myanmar and Bangladesh to build the pipeline. Because Bangladesh and India failed to agree on the terms, this deal failed.[46] At the time, Myanmar government sold the natural gas in Rakhine state to Chinese companies. Later, India put forward an alternate proposal to build a pipeline from Myanmar through the northeastern states to West Bengal. However, the pipeline was economically unfeasible and had to be shelved in July 2009.[47] India is hopeful that it will get access to the bountiful gas reserve of Myanmar. India's more recent blueprint hopes to construct a 1575 km ambitious gas pipeline passing through Sittwe-Aizwal-Silchar-Guwahati-Siliguri-Gaya.[48] India proposed to transport the gas by ship as liquefied natural gas from Myanmar until the pipeline is constructed. The Myanmar government has sold a 30 per cent stake to two Indian companies in the exploration of and production from the offshore natural gas fields near Sittwe.[49]

### *Military Cooperation*

India and Myanmar have developed a defence cooperation mechanism out of their national security concerns and mutual interests. In this regard, India has also improved its relationship with *Tatmadaw*. India provided uniforms and battlefield training to Myanmar armed forces, leased the military a helicopter squadron, and offered its services to maintain Myanmar's equipment.[50] From 2003 to 2004 several port calls and joint naval manoeuvres were organised.[51] From time to time Myanmar's air force officers have received military training from India.[52] Even more, as part of their official duties, the Indo-Tibetan Border Police watches the Myanmar border area.[53]

India is known for providing military hardware to Myanmar. In 2007, quartermaster general, Lieutenant General Thiha Thura Tin Aung Myint Oo of Myanmar visited India with an urge to buy a number of military hardware from the chief of the Indian army. He wanted to buy infantry weapons and ammunition. In return, he offered to help diffuse Indian insurgents. India sells only small arms to Myanmar like assault rifles, machine guns, and side arms. In addition to that, India has trained Myanmar military officers on medical, air force, and naval fields. India has also sold rocket launchers, radar and engineering equipment, night vision systems, and torpedoes to Myanmar.[54]

## CAN CHINA AND INDIA COEXIST IN RAKHINE?

To sum up, China and India have parallel objectives in Myanmar: protection of national interests, particularly ensuring their geopolitical and geo-economic security. Thus, there are obvious reasons for these two countries to be competing against each other.

The brutal crackdown on the Rakhine *Rohingya* has angered Bangladesh and several other neighbouring countries. Bangladesh Prime Minister Sheikh Hasina initially reacted furiously, raising fears of a military confrontation. She nevertheless decided to de-escalate tensions and seek a peaceful resolution. Several diplomatic missions were sent from Dhaka to Naypyidaw since the violence erupted in August 2017. She is also keen to prevent a trans-border extremist nexus from taking hold and has offered support to the *Tatmadaw* to take on the Arakan Rohingya Salvation Army (ARSA), responsible for the August 25, 2017 attack on 30 police outposts in Rakhine.[55]

India too is deeply concerned about the extremist nexus. The Indian intelligence agency has reported close connections between the ARSA, Bangladesh's *Juama'atul Mujahideen* (JMB), and the Indian *Mujahideen*. According to their report, all these groups receive the backing from Pakistan's *Lashkar-e-Taiba*, allegedly responsible for the 2008 terror strike on Mumbai.[56] India's supposition became quite obvious when the Indian Ministry of External Affairs made a statement the day after the ARSA attack: "We stand by Myanmar in the hour of its crisis, we strongly condemn the terrorist attack on August 24–25 and condole the death of policemen and soldiers, we will back Myanmar in its fight."[57] The Indian statement failed to mention Myanmar military's genocidal retaliatory attacks on the *Rohingya* community. To show further support to Myanmar, India threatened to expel nearly 40,000 *Rohingya* migrants settled illegally in the country, among which 16,500 registered with the United Nations High Commissioner for Refugees (UNHCR).[58] India's action drew sharp UN criticism.

Like India, China also extended "strong support" for Myanmar on the Rakhine issue. At a welcoming ceremony marking the 68th anniversary of the founding of People's Republic of China, Chinese ambassador to Myanmar, Hong Liang, stated the "hope that the international community will create a good external environment so that Myanmar can solve its problems properly."[59] His comment was before an open discussion on Rakhine at the UN Security Council.

Scholars agree that enormous Chinese and Indian infrastructure projects in the Rakhine are the key reasons behind their staunch backing for Myanmar, even after *Tatmadaw*'s ruthless ethnic cleansing of *Rohingyas*. Although none of these projects is in the troubled and predominantly Muslim-majority Northern Rakhine, the threat of the spreading of violent extremism to the other parts of Rakhine, where Beijing and New Delhi invested heavily, is a source of great concern for them. The former deputy chief of India's Defence Intelligence Agency Major General Gaganjit Singh raised some pertinent questions: "What if ARSA terrorists attack an Indian ship on the Kaladan River or try blowing up parts of the Yunnan-Kyauk Phyu oil-gas pipeline … Such scenarios cannot be discounted."[60]

The China-India competition in Myanmar shows China ahead of India, but with India trying to reduce that dependence on China through Naypyidaw's policy. Myanmar faces not only a diplomatic challenge but also a domestic plight, given its not so stable political and social conditions: it cannot rely on any one country too much for political and economic sustenance, thus possibly threatening Chinese and Indian interests. Ignoring the ceasefire agreement, Myanmar has continued to dominate and brutalise the ethnic groups, since issues like Rakhine and the *Rohingyas* are a matter of great concern for other parties as well.

Rakhine is only about 150 km away from Chattogram in Bangladesh. The economic and geographic maps of this region are changing rapidly. The continued focus on the Rakhine issue by both India and China, and a policy shift by them from incorporating Bangladesh to negating Bangladesh in Myanmar, reduces the geographic and strategic importance of Bangladesh to them.

During the September 2017 BRICS (Brazil-Russia-India-China-South Africa) Summit, Indian Prime Minister met the Chinese president, and both decided to avoid confrontation and agreed on taking steps towards de-escalating tensions by pulling troops from the Bhutanese territory. They also agreed to work on gaining regional stability, especially on the ongoing *Rohingya* issue.[61]

According to Binod Mishra, head of the Centre for Studies in International Relations and Development in India: "both India and China engage the Burmese military as much as the civilian government because the country is key to India's 'Act East' policy and China's 'One Belt and Road Initiative.'"[62]

Though both India and China seek a truce, they also compete with each other for power and influence. In an interview, Indian Ambassador Vikram Misri said their approach to the development of Myanmar to be different from "others," indicating China, though not saying the word out loud. "India wants to create public assets in Myanmar and hand over to local authorities. We finance projects like Kaladan mostly by grants and some concessional financing, but we ensure that these never become a burden."[63]

Myanmar has other strategic options as well if either India or China tries to influence it too much. It can adopt a balance-of-power strategy and can invite more countries, like Russia or the United States, to participate in its policymaking. Such moves were observed when Myanmar decided to purchase twenty MiG-29s in 2001 and its recent decision to purchase six advanced Sukhoi-30 multirole fighter jets from Russia in January 2018.[64]

Though there is competition for influence between India and China, they could form an alliance if they reach a point when they have to compete with other important national interests.

## Conclusions

Because of its great geographic and strategic importance, both India and China have made significant efforts to improve their relations with Myanmar. Both achieved different results due to their different political policies.

China's and India's competition for influence on Myanmar's foreign policy and accessing control of resources in the country seem set to grow in the foreseeable future. At present, China is enjoying a privileged position in Myanmar, but India's aggressive political and economic agendas in the country may challenge China's influence on some specific areas in the near future. Myanmar, on the other hand, will be the biggest winner as it will try to attain the maximum benefits out of its growing friendly relations with both India and China.

It could be concluded that both India and China will continue to keep their influence on Myanmar even though Myanmar faces a number of challenges such as the ongoing *Rohingya* issue and international denouncements.

## Notes

1. OCHA's latest report indicates that nearly 646,000 Rohingyas entered Bangladesh since August 25. UNOCHA, "Rohingya Refugee Crisis," from: http://interactive.unocha.org/emergency/2017_rohingya/ last consulted 11 December 2018.
2. Islam, "Dragon meets elephant: China and India's stakes in Myanmar," *The Daily Star*, October 12, 2017, from: http://www.thedailystar.net/opinion/perspective/mayanmar-rohingya-refugee-crisis-dragon-meets-elephant-myanmar-1475020, last consulted January 25, 2018.
3. Li Chenyang, "China-Myanmar Relations since 1988," *Harmony and Development: ASEAN China Relations* (Singapore: World Scientific, 2007): 49–64.
4. Yuh-Ming Tsai, "Breakout: China Foreign Policy toward Myanmar," *Feng Chia Journal of Humanities* 8 (2010): 302–325.
5. Shahidul Islam, op. cit.
6. Wai Moe, "China signs Burmese gas deal for 30-year supply," *The Irrawaddy*, December 26, 2008, from: http://www2.irrawaddy.com/article.php?art_id=14849, last consulted February 10, 2018.
7. Wu Hongying, "CPPCC thematic consultations for the proposal of building the Indian Ocean channel," 21st-Century Business Herald, March 10, 2009, from: www1.21cbh.com/HTML/2009-3-10/HTML_SKBAI7DM7BKJ.html, last consulted February 10, 2018.
8. Islam, op. cit.
9. Xiangming Chen, "China's Key Cities: From Local Places to Global Players," *East by Southeast*, January 7, 2016, from: http://www.eastbysoutheast.com/tag/ruili/, last consulted February 18, 2018.
10. Ibid.; and Islam, op. cit.
11. Jürgen Haacke, *Myanmar's Foreign Policy: Domestic Influences and International Implications* (New York: Routledge, 2006).
12. Irina Slav, "China's oil import dependency deepens," The Oil Price.com, Washington DC, January 13, 2017, from: https://oilprice.com/Latest-Energy-News/World-News/Chinas-Oil-Import-Dependency-Deepens.html, last consulted February 22, 2018.
13. Zhao Hong, "India and China: Rivals or partners in Southeast Asia?" *Contemporary Southeast Asia* 29, no. 1 (2007): 121–143.

CHINA, INDIA, AND MYANMAR: PLAYING ROHINGYA ROULETTE? 93

94    H. A. TAUFIQ

CHINA, INDIA, AND MYANMAR: PLAYING ROHINGYA ROULETTE? 95

96    H. A. TAUFIQ56. Bhaumik, op. cit.
57. Ibid.
58. Human Rights Watch. "India: Don't forcibly return Rohingya refugees." August 17, 2017, from: https://www.hrw.org/news/2017/08/17/india-dont-forcibly-return-rohingya-refugees, last consulted 19 February 2018.
59. Bhaumik, op. cit.
60. Ibid.
61. Ibid.
62. Ibid.
63. Ibid.
64. The Myanmar Times. "Russia to sell six fighter jets to Myanmar," January 23, 2018, from: https://www.mmtimes.com/news/russia-sell-six-fighter-jets-myanmar.html, last consulted 22 February 2018; and Atul Bharadwaj, "Myanmar and MiG-29s," *Institute of Peace and Conflict Studies*, 2002. http://www.ipcs.org/article/military-and-defence/myanmar-and-mig-29s-751.html, last consulted 23 February 2018.

## Bibliography

Bharadwaj, Atul. 2002. *Myanmar and MiG-29s*. Institute of Peace and Conflict Studies. From: http://www.ipcs.org/article/military-and-defence/myanmar-and-mig-29s-751.html

Bhaumik, Subir. 2017. *Why do China, India back Myanmar over Rohingya crisis?* Prod. The Week in Asia. October 18. From: http://www.scmp.com/week-asia/geopolitics/article/2115839/why-do-china-india-back-myanmar-over-rohingya-crisis

Chen, Xiangming. 2016. "China's Key Cities: From local places to global players." *East by Southeast*. January 7. From: http://www.eastbysoutheast.com/tag/ruili/

Chenyang, Li. 2007. "China-Myanmar relations since 1988." In *Harmony and Development: ASEAN-China Relations*. Ed., Lai Hongyi and Lim Tin Seng, 49–64. Singapore: World Scientific.

Chenyang, Li. 2010. "The Policies of China and India toward Myanmar." In *Myanmar/Burma: Inside Challenges, Outside Interests*. Ed., Lex Rieffel, 114–115. Washington DC: Konrad Adenauer Foundation, Brookings Institution Press.

U.S. Energy Information Administration. 2016. "Country Analysis Brief: India (2014, 29)." *S&P Global*. June 14. From: http://www.ieee.es/Galerias/fichero/OtrasPublicaciones/Internacional/2016/EIA_Country_Analysis_Brief_India_14jun2016.pdf. Accessed February 24, 2018.

CHINA, INDIA, AND MYANMAR: PLAYING ROHINGYA ROULETTE?   99Slav, Irina. 2017. *China's Oil Import Dependency Deepens*. Washington DC, January 13. Accessed February 22, 2018. https://oilprice.com/Latest-Energy-News/World-News/Chinas-Oil-Import-Dependency-Deepens.html
TNN. 2006. "Gail picks up 30% in Myanmar block." *The Economic Times*. December 9. https://economictimes.indiatimes.com/industry/energy/oil-gas/gail-picks-up-30-in-myanmar-block/articleshow/748562.cms. Accessed February 17, 2018.
———. 2003. "Indo-Thai-Myanmar Highway." *The Times of India*. December 24. https://timesofindia.indiatimes.com/india/Indo-Thai-Myanmar-highway/articleshow/378309.cms. Accessed January 15, 2018.
Thiha, Amara. 2017. *The Bumpy Relationship Between India and Myanmar: Delhi-Naypyidaw relations may fall short, but India has made progress in some areas*. August 25. https://thediplomat.com/2017/08/the-bumpy-relationship-between-india-and-myanmar/
Tsai, Yuh-Ming. 2010. "Breakout: China foreign policy toward Myanmar." *Feng Chia Journal of Humanities* (8): 302–325.
UNOCHA. 2018. "Rohingya Refugee Crisis." http://interactive.unocha.org/emergency/2017_rohingya/. Accessed December 11, 2018.View publication stats